\documentclass[11pt,superscriptaddress,aps,prd,preprint]{revtex4}

\everymath{\displaystyle}
\usepackage{graphicx}

\usepackage{amsmath,amssymb,mathrsfs}

\newcommand{\bea}{\begin{eqnarray}}
\newcommand{\eea}{\end{eqnarray}}

\begin{document}

\title{On the Temperature of Gravitation in the de Sitter Space-Time}

\author{A. F. Santos}\email[]{alesandroferreira@fisica.ufmt.br}
\affiliation{Instituto de F\'{\i}sica, Universidade Federal de Mato Grosso,\\
78060-900, Cuiab\'{a}, Mato Grosso, Brazil}

\author{S. C. Ulhoa}\email[]{sc.ulhoa@gmail.com}
\affiliation{International Center of Physics, Instituto de F\'isica, Universidade de Bras\'ilia, 70910-900, Bras\'ilia, DF, Brazil} \affiliation{Canadian Quantum Research Center,\\ 
204-3002 32 Ave Vernon, BC V1T 2L7  Canada} 

\author{T. F. Furtado}
\email{tarciofla@hotmail.com}
\affiliation{Instituto de F\'isica, Universidade de Bras\'ilia, 70910-900, Bras\'ilia, DF, Brazil}

\author{Faqir C. Khanna\footnote{Professor Emeritus - Physics Department, Theoretical Physics Institute, University of Alberta\\
Edmonton, Alberta, Canada}}\email[]{khannaf@uvic.ca}
\affiliation{Department of Physics and Astronomy, University of Victoria,\\
3800 Finnerty Road Victoria, BC, Canada}

\begin{abstract}

The temperature associated with gravitation as presented in the Unruh effect and Hawking temperature serves to link different areas of physics, such as gravity, statistical mechanics, and quantum physics. In this paper, teleparallel gravity is considered to study temperature effects on the de Sitter space-time. The effects of temperature are introduced using the Thermo Field Dynamics (TFD) formalism. The gravitational Stefan-Boltzmann law is obtained. Then the temperature of gravitation in the de Sitter space-time is calculated. Here an Unruh-type effect is discussed. This effect relates the temperature and the acceleration of a particle in the de Sitter space-time. The gravitational Casimir effect is calculated. The result shows that there is a transition between an attractive and a repulsive Casimir effect.

\end{abstract}
\maketitle

\date{\today}
\section{Introduction} \label{sec.1}

At the present the big challenge in physics is to unify quantum mechanics and general relativity into a comprehensive theory of quantum gravity. Although a quantum version of gravity has been under construction for a long time, there is not a consistent quantum theory for gravitation. However, there are some studies that connect gravitational effects and quantum mechanics, such as the Hawking radiation \cite{penrose, Bekenstein, hawking}. It is a radiation that coming from a black hole. This radiation has a thermal spectrum and an absolute temperature, known as the Hawking temperature. An effect closely related to the black hole radiation is the Unruh effect \cite{Fulling, Davies, Unruh}. The Unruh and Hawking effects serve to link the three main branches of modern physics, i.e., statistical physics, gravitation and quantum physics. Together they are widely regarded as forming a valuable sign to the search for a quantum theory of gravity. The Unruh effect implies that inertial observers see no particles when fields are in their inertial vacuum state. But uniformly accelerated observers (also called Rindler observers) see a thermal state of particles at a temperature $T$ which is proportional to their proper acceleration $a$, i.e., $T=a/2\pi$. For a review on the Unruh effect, see \cite{Crispino}. This phenomenon was initially discovered for flat space-time, and then has been extended to space-times with curvature and cosmological horizon. In curved space-time, there are relations between Unruh temperature and cosmological constant \cite{Gib, Crispino}. For a better understanding of the Unruh effect in curved space-time, Unruh and Dewitt \cite{Unruh, Dewitt, Birrel} developed a concept of particle detector to examine the Unruh radiation. Several investigations have concluded that accelerated detectors get hotter. So, a natural question arises: is it a universal phenomenon? The answer is no. It has been shown that a detector coupled to the scalar field vacuum for finite timescales has a temperature that decreases with acceleration, in certain regimes \cite{Anti}. It is called the anti-Unruh effect, i.e., a uniformly accelerated particle detector coupled to the vacuum can cool down as its acceleration increases. 

In this paper, the main objective is to investigate the temperature effect of gravitation in the de Sitter space-time. This space-time is the maximally symmetric solution of the Einstein equations with a positive cosmological constant. There are two different ways of introducing the effects of temperature into a field theory. The first is the Matsubara formalism \cite{Matsubara}. In this approach, the Wick rotation is performed and  this leads to a substitution of time, $t$, by a complex time, $i\tau$. The second approach is the real-time formalism. There are two distinct, but equivalent approaches to real-time formalism. The closed time path formalism \cite{Schwinger} and the Thermo Field Dynamics (TFD) formalism \cite{Umezawa1, Umezawa2, Khanna0, Umezawa22, Khanna1, Khanna2}. TFD is a real-time formalism which consists of two elements: (i) the doubling of the original Hilbert space, composed of the original and tilde (dual) space and (ii) the Bogoliubov transformation. The Bogoliubov transformation is a rotation between two spaces, original and dual. The application of TFD in gravitation essentially produces a theory of quantum gravity.  Like any quantum structure there are the observables of the theory which are the average quantities.  There are approaches that focus more on quantum operators \cite{Acquaviva} and there are those that are concerned with observables that have more relevant experimental implications.  In that article this last line is followed and for that reason effects such as the Stefan-Boltzmann and Casimir law are investigated. The TFD formalism is used here to investigate the effects of temperature on teleparallel gravity in the de Sitter space-time. Such a formalism is entirely equivalent to Matsubara approach \cite{abdala}.

The main ideas of the teleparallel gravity came up with Einstein using his attempts to unify gravity and electromagnetism \cite{Einstein}. In this paper the Teleparallel Equivalent of General Relativity (TEGR) is considered. A review of this gravitational theory has been carried out \cite{MalufRev}. In this framework the torsion is responsible for the dynamics of the space-time. TEGR is a gravitational theory where the fundamental variables are the tetrad fields. This description is dynamically equivalent to GR. The main advantage of TEGR when compared to general relativity is the fact that there is a well-defined expression for the gravitational energy-momentum tensor. The total gravitational energy and the gravitational energy density of the de Sitter space-time, using the definition of localized gravitational energy that naturally arises in the framework of TEGR, has been calculated \cite{Maluf}. Furthermore, although TEGR and general relativity are equivalent, the definition of graviton propagator in a linearized approach is quite different \cite{flatTFD}.

This paper is organized as follows. In section II, a brief introduction to the TFD formalism is presented. In section III, the teleparallel gravity is introduced. The gravitational energy-momentum tensor is defined. In section IV, the de Sitter space-time is presented where the acceleration of a particle is discussed. Then the gravitational Stefan-Boltzmann law in the de Sitter space-time is obtained. Using the Maxwell relations and the gravitational entropy, a relationship between temperature and particle acceleration is constructed. It is an Unruh-type effect. In addition, the gravitational Casimir effect at zero and finite temperature are calculated. The Casimir effect was first studied by Casimir in 1948 \cite{Casimir}. It describes the interaction between two parallel conducting plates. The plates modify the quantum vacuum and as a result the plates are attracted towards each other. Although at the first moment, this effect was analyzed for the electromagnetic field, it can be defined for any quantum field also. Here the gravitational Casimir effect at zero temperature is calculated. The acceleration of the particle in the de Sitter space-time changes this effect. A transition between an attractive and a repulsive Casimir effect is obtained. Furthermore, the Casimir effect at finite temperature is investigated. In section V, some concluding remarks are presented.

\section{Thermo Field Dynamics (TFD)} \label{sec.2}

The TFD is a quantum field theory at finite temperature. It is defined to be real-time finite temperature formalism \cite{Umezawa1, Umezawa2, Khanna0, Umezawa22, Khanna1, Khanna2}. Unlike imaginary formalism \cite{Matsubara}, it allows a study of both effects, temperature and time evolution of the system. The main objective of this formalism is to write the vacuum expectation value of an arbitrary operator $A$ equal to the statistical average, that is,
\bea
\langle A \rangle=\langle 0(\beta)| A|0(\beta) \rangle,
\eea
where  $|0(\beta) \rangle$ is the thermal vacuum state with $\beta=\frac{1}{k_BT}$, $T$ being the temperature and $k_B$ the Boltzmann constant. This thermal state is developed from two fundamental elements: the doubling of the original Hilbert space and the Bogoliubov transformation. This doubling consist of ${\cal S}_T={\cal S}\otimes \tilde{\cal S}$, where ${\cal S}$ is the Hilbert space and $\tilde{\cal S}$ is the dual (tilde) space. The Bogoliubov transformation introduces thermal effects through a rotation between tilde ($\tilde{\cal S}$) and non-tilde (${\cal S}$) operators. The Bogoliubov transformation is defined as
\bea
\left( \begin{array}{cc} {\cal O}(k, \alpha)  \\\xi \tilde {\cal O}^\dagger(k, \alpha) \end{array} \right)={\cal B}(\alpha)\left( \begin{array}{cc} {\cal O}(k)  \\ \xi\tilde {\cal O}^\dagger(k) \end{array} \right),
\eea
where $\xi = -1 (+1)$ for bosons (fermions), ${\cal O}$ is an arbitrary operator,  the $\alpha$ parameter is called the compactification parameter defined by $\alpha=(\alpha_0,\alpha_1,\cdots\alpha_{D-1})$ and ${\cal B}(\alpha)$ is 
\bea
{\cal B}(\alpha)=\left( \begin{array}{cc} u(\alpha) & -w(\alpha) \\
\xi w(\alpha) & u(\alpha) \end{array} \right),
\eea
with $u^2(\alpha)+\xi w^2(\alpha)=1$. Furthermore, these quantities $u(\alpha)$ and $w(\alpha)$ are related to the Bose or Fermi distributions. 

It is interesting to note that, using the TFD formalism, all propagators can be expressed in terms of the compactification parameter $\alpha$. As a simple example, the scalar field propagator is considered. It is given as
\bea
G_0^{(AB)}(x-x';\alpha)=i\langle 0,\tilde{0}| \tau[\phi^A(x;\alpha)\phi^B(x';\alpha)]| 0,\tilde{0}\rangle,
\eea
where $A\, \mathrm{and} \,B=1,2$ and $\tau$ is the time ordering operator. The Bogoliubov transformation is used to write the scalar field as
\bea
\phi(x;\alpha)&=&{\cal B}(\alpha)\phi(x){\cal B}^{-1}(\alpha).
\eea
By taking $|0(\alpha)\rangle={\cal B}(\alpha)|0,\tilde{0}\rangle$, as the thermal vacuum, the scalar field propagator becomes
\bea
G_0^{(AB)}(x-x';\alpha)&=&i\langle 0(\alpha)| \tau[\phi^A(x)\phi^B(x')]| 0(\alpha)\rangle,\nonumber\\
&=&i\int \frac{d^4k}{(2\pi)^4}e^{-ik(x-x')}G_0^{(AB)}(k;\alpha),
\eea
where
\bea
G_0^{(AB)}(k;\alpha)={\cal B}^{-1}(\alpha)G_0^{(AB)}(k){\cal B}(\alpha),
\eea
with
\bea
G_0^{(AB)}(k)=\left( \begin{array}{cc} G_0(k) & 0 \\
0 & \xi G^*_0(k) \end{array} \right),
\eea
and
\bea
G_0(k)=\frac{1}{k^2-m^2+i\epsilon},
\eea
where $m$ is the scalar field mass. Then the Green function is written as
\bea
G_0^{(11)}(k;\alpha)=G_0(k)+\xi w^2(k;\alpha)[G^*_0(k)-G_0(k)].
\eea
Note that only the component with $A=B=1$ has been considered. This is due to the fact that physical quantities are given by the non-tilde variables. The quantity $w^2(k;\alpha)$ is the generalized Bogoliubov transformation \cite{GBT} which is defined as
\bea
w^2(k;\alpha)=\sum_{s=1}^d\sum_{\lbrace\sigma_s\rbrace}2^{s-1}\sum_{l_{\sigma_1},...,l_{\sigma_s}=1}^\infty(-\xi)^{s+\sum_{r=1}^sl_{\sigma_r}}\,\exp\left[{-\sum_{j=1}^s\alpha_{\sigma_j} l_{\sigma_j} k^{\sigma_j}}\right],\label{BT}
\eea
where $\lbrace\sigma_s\rbrace$ denotes the set of all combinations with $s$ elements and $k$ is the 4-momentum.

\section{Teleparallel Gravity} \label{sec.3}

The general relativity is a gravitational theory constructed in Riemann geometry. An alternative theory is the Teleparallel Equivalent of General Relativity (TEGR), that is developed in Weitzenb\"{o}ck geometry. Although these theories are defined in different geometries, they are dynamically equivalent. In TEGR, the field variables are tetrads, rather than the metric tensor. The tetrad theory of gravity is a geometrical framework more general than the Riemannian geometry. In Weitzenb\"{o}ck geometry, it is possible to construct the Cartan connection \cite{cartan}. It is defined as
\bea
\Gamma_{\mu\lambda\nu}=e^{a}\,_{\mu}\partial_{\lambda}e_{a\nu}\,.
\eea
The metric tensor is obtained from the tetrad fields by $g_{\mu\nu}=e^{a}\,_{\mu}e_{a}\,_{\nu}$. The Cartan connection has zero curvature but has a torsion given as
\begin{equation}
T^{a}\,_{\lambda\nu}=\partial_{\lambda} e^{a}\,_{\nu}-\partial_{\nu}
e^{a}\,_{\lambda}\,. \label{4}
\end{equation}
Note that, the Cartan (or Weitzenb\"{o}ck) connection is related to the Christoffel symbols, ${}^0\Gamma_{\mu \lambda\nu}$, as
\begin{equation}
\Gamma_{\mu \lambda\nu}= {}^0\Gamma_{\mu \lambda\nu}+ K_{\mu
\lambda\nu}\,, \label{2}
\end{equation}
where $K_{\mu\lambda\nu}$ is the contortion tensor that is defined as
\begin{eqnarray}
K_{\mu\lambda\nu}&=&\frac{1}{2}(T_{\lambda\mu\nu}+T_{\nu\lambda\mu}+T_{\mu\lambda\nu})\,,\label{3}
\end{eqnarray}
with $T_{\lambda\mu\nu}=e_{\mu a}T^{a}\,_{\lambda\nu}$.

The equivalence between TEGR and the theory of general relativity is guaranteed by the identity
\begin{equation}
eR(e)\equiv -e(\frac{1}{4}T^{abc}T_{abc}+\frac{1}{2}T^{abc}T_{bac}-T^aT_a)+2\partial_\mu(eT^\mu)\,,\label{eq5}
\end{equation}
where $e=det(e^a\,_\mu)$, $R$ is the curvature scalar calculated from the Christoffel symbols and $T^a=e^a\,_\mu T^\nu\,_\nu\,^\mu$. In addition, the Lagrangian density that describes the teleparallel gravity and it is equivalent to the Einstein-Hilbert Lagrangian is
\begin{eqnarray}
{\cal L}(e_{a\mu})&=& -\kappa\,e\,(\frac{1}{4}T^{abc}T_{abc}+
\frac{1}{2} T^{abc}T_{bac} -T^aT_a) -{\cal L}_M\nonumber \\
&\equiv&-\kappa\,e \Sigma^{abc}T_{abc} -{\cal L}_M\;, \label{6}
\end{eqnarray}
with $\kappa=1/(16 \pi)$, ${\cal L}_M$ being the Lagrangian density of matter fields and $\Sigma^{abc}$ is a tensor defined as
\begin{equation}
\Sigma^{abc}=\frac{1}{4} (T^{abc}+T^{bac}-T^{cab}) +\frac{1}{2}(\eta^{ac}T^b-\eta^{ab}T^c)\;. \label{7}
\end{equation}

The field equations  are derived from arbitrary variations of ${\cal L}(e_{a\mu})$  with respect to tetrads. This leads to
\begin{equation}
\partial_\nu\left(e\Sigma^{a\lambda\nu}\right)=\frac{1}{4\kappa}
e\, e^a\,_\mu( t^{\lambda \mu} + T^{\lambda \mu})\;, \label{10}
\end{equation}
where $t^{\lambda \mu}$ is interpreted as the gravitational energy-momentum tensor. It is given as
\begin{equation}
t^{\lambda \mu}=\kappa\left[4\,\Sigma^{bc\lambda}T_{bc}\,^\mu- g^{\lambda
\mu}\, \Sigma^{abc}T_{abc}\right]\,. \label{11}
\end{equation}
From this, a local conservation law is obtained as
\begin{equation}
\partial_\lambda\partial_\nu\left(e\Sigma^{a\lambda\nu}\right)\equiv0\,.\label{12}
\end{equation}
This is a consequence of the skew-symmetric property of $\Sigma^{a\lambda\nu}$.

Using these results, it is possible to write an energy-momentum vector of the gravitational field as
\bea
P^a &=& \int_V d^3x \,e\,e^a\,_\mu(t^{0\mu}+ T^{0\mu})\,.
\eea
It is interesting to note that, such a quantity is invariant under coordinate transformations of the three-dimensional space and Lorentz symmetry, it transforms like a vector as a proper expression of the energy-momentum tensor.

In order to investigate some applications, the graviton propagator for the teleparallel gravity in the weak-field approximation is calculated. Let us consider that the metric tensor is defined as 
\bea
g_{\mu\nu}=\eta_{\mu\nu}+h_{\mu\nu},
\eea
where $\eta_{\mu\nu}$ is the Minkowski metric and $h_{\mu\nu}$ a small perturbation term. Using the Lagrangian (\ref{6}) the graviton propagator is 
\begin{equation}
\langle e_{b\lambda}, e_{d\gamma} \rangle=\Delta_{bd\lambda\gamma} = \frac{\eta_{bd}}{\kappa q^{\lambda} q^{\gamma}}.
\end{equation}
More details about the graviton propagator are given in reference \cite{flatTFD}. The graviton propagator leads to the Green function that reads
\bea
G_0(x,x')=-i\Delta_{bd\lambda\gamma}\,g^{\lambda\gamma}\eta^{bd}.
\eea
Explicitly it is
\begin{equation}
G_0(x,x')= -\frac{i64\pi}{q^{2}}\,,
\end{equation}
with $q=x-x'$, where $x$ and $x'$ are four vectors. With the weak field approximation, the gravitational energy-momentum tensor $t^{\lambda \mu}$ becomes
\begin{eqnarray}
t^{\lambda\mu}(x) &=& \kappa\Bigl[g^{\mu\alpha}\partial^{\gamma}e^{b\lambda}\partial_{\gamma}e_{b\alpha} - g^{\mu\gamma}\partial^{\alpha}e^{b\lambda}\partial_{\gamma}e_{b\alpha} - g^{\mu\alpha}(\partial^{\lambda}e^{b\gamma}\partial_{\gamma}e_{b\alpha} - \partial^{\lambda}e^{b\gamma}\partial_{\alpha}e_{b\gamma})\nonumber\\
        & &-2g^{\lambda\mu}\partial^{\gamma}e^{b\alpha}(\partial_{\gamma}e_{b\alpha}-\partial_{\alpha}e_{b\gamma})\Bigl]\,.
\end{eqnarray}

To study some applications using the TFD formalism, the vacuum expectation value of the energy-momentum tensor is calculated. In additon, the energy-momentum tensor is written so as to avoid a product of field operators at the same space-time point. Then
\bea
\langle t^{\lambda\mu}(x)\rangle&=& \langle 0|t^{\lambda\mu}(x)|0\rangle,\nonumber\\
&=& \lim_{x^\mu\rightarrow x'^\mu} 4i\kappa\left(-5g^{\lambda\mu}\partial'^{\gamma}\partial_{\gamma} +2g^{\mu\alpha}\partial'^{\lambda}\partial_{\alpha}\right)G_{0}(x-x')\,,\label{em}
\eea
where $\langle e_{c}^{\,\,\,\lambda}(x), e_{b\alpha}(x') \rangle = i\eta_{cb}\,\delta^{\lambda}_{\alpha}\,G_{0}(x-x')$. 

Using ingredients of the TFD formalism, the vacuum expectation value of the energy-momentum tensor becomes
\bea
\langle t^{\lambda\mu(AB)}(x;\alpha)\rangle=\lim_{x\rightarrow x'} 4i\kappa\left(-5g^{\lambda\mu}\partial'^{\gamma}\partial_{\gamma} +2g^{\mu\alpha}\partial'^{\lambda}\partial_{\alpha}\right)G_{0}^{(AB)}(x-x';\alpha).
\eea
In order to obtain a finite quantity, the Casimir prescription is used, i.e.,
\bea
{\cal T}^{\lambda\mu (AB)}(x;\alpha)=\langle t^{\lambda\mu(AB)}(x;\alpha)\rangle-\langle t^{\lambda\mu(AB)}(x)\rangle\,.
\eea
Then the energy-momentum tensor becomes
\bea
{\cal T}^{\lambda\mu (AB)}(x;\alpha)=\lim_{x\rightarrow x'} 4i\kappa\left(-5g^{\lambda\mu}\partial'^{\gamma}\partial_{\gamma} +2g^{\mu\alpha}\partial'^{\lambda}\partial_{\alpha}\right)\overline{G}_{0}^{(AB)}(x-x';\alpha),\label{EM}
\eea
where
\bea
\overline{G}_0^{(AB)}(x-x';\alpha)=G_0^{(AB)}(x-x';\alpha)-G_0^{(AB)}(x-x').
\eea

In the next section, using the energy-momentum tensor (\ref{EM}), some applications at finite temperature in the de Sitter space-time are calculated.

\section{Stefan-Boltzmann law and Casimir effect in de Sitter Space-Time } \label{sec.4}

In this section, the de Sitter space-time is considered. The total gravitational energy in the framework of TEGR of the de Sitter space-time has been studied in detail \cite{Maluf}. In this background, the Stefan-Boltzmann law and the Casimir effect at finite temperature are calculated.

The metric that describes the de Sitter universe is
\begin{equation}
ds^{2} = -\left( 1-\frac{r^{2}}{R^{2}} \right)dt^{2} + \left( 1-\frac{r^{2}}{R^{2}} \right)^{-1}dr^{2} + r^{2}d\Omega^{2},\label{Metric}
\end{equation}
with $d\Omega^{2} = d\theta^{2} + \sin^{2}\theta d\phi^{2}$. The de Sitter space-time is a solution to the gravitational field equations without any source and with  a cosmological constant $\Lambda$. The de Sitter radius $R$ is related to  the cosmological constant  by
\bea
R = \sqrt{\frac{3}{\Lambda}}.
\eea

A direct inspection of the line element (\ref{Metric}) shows that the $g_{00}$ component can be written in the form 
\bea
g_{00}=1+2\Phi
\eea
where $\Phi$ is given as
\bea
\Phi=-\frac{1}{6}\Lambda r^2.
\eea
In order to be clear, it is assumed that as a source of matter is, $m\neq 0$. In this case, the metric describes the Schwarzschild-de Sitter solution as
\begin{equation}
ds^{2} = -\left( 1-\frac{2mG}{r}-\frac{r^{2}}{R^{2}} \right)dt^{2} + \left( 1-\frac{2mG}{r}-\frac{r^{2}}{R^{2}} \right)^{-1}dr^{2} + r^{2}d\Omega^{2}.\label{Metric2}
\end{equation}
This leads to
\bea
\Phi=-\frac{mG}{r}-\frac{1}{6}\Lambda r^2.
\eea
Therefore, $\Phi$ is the gravitational potential. Then this potential implies the motion of a test particle along the geodesics of the metric (\ref{Metric2}). In this case, the acceleration due to the potential (using that $\vec{a}=-\vec{\nabla}\Phi$) is
\bea
a=-\frac{mG}{r^2}+\frac{1}{3}\Lambda r.
\eea
It is to be noted that, even in the case of the de Sitter solution that occurs in the absence of a mass (it is the case considered hereafter) a test particle would be subject to a radial acceleration
\bea
a=\frac{1}{3}\Lambda r.
\eea
Therefore the acceleration increases with the distance. This makes the gravitational field more intense at points far from the origin \cite{Maluf}.

Now these results are used to calculate the gravitational Stefan-Boltzmann law and the Casimir effect at finite temperature. A topological quantum field theory is considered. Here a topology $\Gamma_D^d=(\mathbb{S}^1)^d\times \mathbb{R}^{D-d}$ with $1\leq d \leq D$ is used. In this notation, $D$ are the space-time dimensions and $d$ is the number of compactified dimensions. Then any set of dimensions of the manifold $\mathbb{R}^{D}$ can be compactified, where the circumference of the $nth$ $\mathbb{S}^1$ is specified by $\alpha_n$. For example, the effect of temperature is described by the choice $\alpha_0\equiv\beta$ and $\alpha_1,\cdots\alpha_{D-1}=0$.

\subsection{Gravitational Stefan-Boltzmann Law }

In order to calculate the gravitational Stefan-Boltzmann Law, the topology $\Gamma_4^1=\mathbb{S}^1\times\mathbb{R}^{3}$ is considered, with $\alpha=(\beta,0,0,0)$. Then the time-axis is compactified in $\mathbb{S}^1$, with circumference $\beta$. The generalized Bogoliubov transformation becomes
\bea
w^2(\beta)=\sum_{j_0=1}^\infty e^{-\beta k^0 j_0}.
\eea
The Green function is
\bea
\overline{G}_0^{(11)}(x-x';\beta)&=&2\sum_{j_0=1}^\infty G_0^{(11)}\left(x- x'-i\beta j_0 n_0\right),\label{1GF}
\eea
where $n_0=(1,0,0,0)$. Then the energy-momentum tensor (\ref{EM}) becomes
\bea
{\cal T}^{00 (11)}(x;\beta) &=& 8ik\lim_{x\rightarrow x'}\sum_{j_0=1}^\infty\Bigg\{ -3\left( 1+\frac{2r^{2}}{R^{2}} \right)\partial'_{0}\partial_{0}+5\partial'_{1}\partial_{1} + \frac{5}{r^{2}}\left( 1+\frac{r^{2}}{R^{2}} \right)\nonumber\\
&\times&\Bigg[\partial'_{2}\partial_{2} + \frac{1}{\sin^{2}\theta}\partial'_{3}\partial_{3} \Bigg] \Bigg\}G_0^{(11)}\left(x- x'-i\beta j_0 n_0\right),
\eea
where
\bea
G_0^{(11)}\left(x- x'-i\beta j_0 n_0\right)=-\frac{i64\pi}{(x- x'-i\beta j_0 n_0)^{2}}\label{G}
\eea
with
\begin{eqnarray}
(x- x'-i\beta j_0 n_0)^{2} &=& g_{\mu\nu}(x^\mu- x'^\mu-i\beta j_0 n_0^\mu) (x^\nu- x'^\nu-i\beta j_0 n_0^\nu)\nonumber\\
&-&\left( 1-\frac{r^{2}}{R^{2}} \right)(t-t'-i\beta j_0)^{2} +\left( 1+\frac{r^{2}}{R^{2}} \right)(r-r')^{2} \nonumber\\
&+& r^{2}(\theta-\theta')^{2} + r^{2}\sin^{2}(\phi-\phi')^{2}\,,\label{G1}
\end{eqnarray}
which represents the topological nature of Bogoliubov transformation in TFD. Using these results, the $00$ component of the energy-momentum tensor is 
\begin{equation}
	{\cal T}^{00 (11)}(\beta)=E= \dfrac{32\pi^{4}}{15}\left( 1 + \dfrac{3r^{2}}{R^{2}}\right)T^{4}.\label{energy}
\end{equation}
This is the gravitational Stefan-Boltzmann law in the de Sitter space-time. In addition, taking the component $\lambda=\mu=1$, the gravitational pressure is 
\bea
{\cal T}^{11 (11)}(x;\beta) &=&P= 8ik\lim_{x\rightarrow x'}\sum_{j_0=1}^\infty\Bigg\{ 5\partial'_{0}\partial_{0}-3 \left( 1-\frac{2r^{2}}{R^{2}} \right)\partial'_{1}\partial_{1} - \frac{5}{r^{2}}\left( 1-\frac{r^{2}}{R^{2}} \right)\nonumber\\
&\times&\Bigg[\partial'_{2}\partial_{2} + \frac{1}{\sin^{2}\theta}\partial'_{3}\partial_{3} \Bigg] \Bigg\}G_0^{(11)}\left(x- x'-i\beta j_0 n_0\right).
\eea
Using eqs. (\ref{G}) and (\ref{G1}) the pressure in the de Sitter space-time becomes
\begin{equation}
	P = \dfrac{32\pi^{4}}{45}\left( 1 + \dfrac{r^{2}}{R^{2}}\right)T^{4}.
\end{equation}
From the Maxwell relations, 
\begin{equation}
	P = -\left( \dfrac{\partial F}{\partial V} \right) \quad \quad \quad \mathrm{and} \quad \quad \quad S = -\left( \dfrac{\partial F}{\partial T}\right)
\end{equation}
where $F$ is the free energy, $V$ is the volume and $T$ is the temperature,  it leads to
\begin{equation}
	S = \dfrac{128}{45}\pi^{4}T^{3}\int \left( 1 + \dfrac{r^{2}a}{r} \right)dv
\end{equation}
with $S$ being the entropy associated to the gravity in the de Sitter space-time. Performing an integration, the gravitational entropy is 
\begin{equation}
	\delta S = \dfrac{8}{45}\pi^{3}T^{3}aA^{2},
\end{equation}
where $\delta S = S -S_{0}$ with $S_{0}=\frac{128}{45}\pi^4T^3 V$. By taking $\delta S = A^2$, the temperature is 
\begin{equation}
 T = \sqrt[3]{\dfrac{45}{8\pi^{3}a}}\label{T}.
\end{equation}
This is the gravitational temperature in the de Sitter space-time. Although different, it is a temperature of the Unruh-type since it relates temperature and acceleration. Indeed, our result shows that the temperature of the particle cools down as its acceleration increases, this is known as anti-Unruh effect~\cite{Garay}. 

It is worth noting that we are essentially dealing with a gas of gravitons. This is due to the fact that the operators used in the description of the TFD, create or destroy particles in the field, that is, gravitons.  From this structure, measurable average values are obtained such as energy, pressure and temperature. The energy of this system, given by equation (\ref{energy}), depends directly on the redshift, z.  This can be seen through the relationship between acceleration and z, which for $m=0$ reads
\begin{equation}
a^{2}=\frac{2}{3}\left[\frac{\Lambda z\left(z+2\right)}{\left(z+1\right)^2}\right]\,,
\end{equation}
where $z$ is the redshift. The temperature dependence with the redshift is explicitly given by 
\begin{equation}
T=\sqrt[3]{\frac{45 \sqrt{3}\left(z+1\right)}{8{\pi}^3\sqrt{2\,\Lambda z\left(z+2\right)}}}\,.
\end{equation}
It is interesting to note that when the redshift goes to zero, the acceleration vanishes. In such a limit the temperature goes to infinity. This means that there is a physical limit established by equation (\ref{T}). This limit is a proper prediction of the TFD applied to the gravitational field.

\subsection{Gravitational Casimir effect at zero temperature}

Here the topology $\Gamma_4^1$ with $\alpha=(0,i2d,0,0)$ is used. Then the compactification along the coordinate $r$ is considered. Then the Bogoliubov transformation is 
\bea
w^2(d)=\sum_{l_1=1}^{\infty}e^{-i2d k^1l_1}\label{BT2}
\eea
and the Green function becomes
\bea
\overline{G}_0(x-x';d)=2\sum_{l_1=1}^{\infty}G_0(x-x'-2d l_1n_1)\label{GF2}
\eea
with $n_1=(0,1,0,0)$. Thus the energy-momentum tensor is 
\bea
{\cal T}^{\lambda\mu (AB)}(x;d)=\lim_{x\rightarrow x'}\sum_{l_1=1}^{\infty} 8i\kappa\left(-5g^{\lambda\mu}\partial'^{\gamma}\partial_{\gamma} +2g^{\mu\alpha}\partial'^{\lambda}\partial_{\alpha}\right)G_0(x-x'-2d l_1n_1).
\eea
The Casimir energy and Casimir pressure at zero temperature are calculated by taking the components $00$ and $11$, respectively. Then
\bea
{\cal T}^{00 (11)}(d)&=&-\frac{4\pi^4}{45d^4}\left(1-ar\right),\\
{\cal T}^{11 (11)}(d)&=&-\frac{4\pi^4}{15d^4}\left(1-3ar\right).
\eea
It is important to note that, although the acceleration modifies these quantities, in its absence, the Casimir effect is due to the flat space-time. Furthermore, the Casimir pressure can be attractive or repulsive. Using the acceleration that is given as $a=\frac{1}{3}\Lambda r$, the pressure becomes
\bea
P=-\frac{4\pi^4}{15d^4}\left(1-\Lambda r^2\right),
\eea
where $P={\cal T}^{11 (11)}(d)$. Then there is a critical value $\Lambda=\frac{1}{r^2}$ for the transition from negative to positive values of the pressure. In other words, the Casimir force is attractive for $\Lambda<\frac{1}{r^2}$ and repulsive for $\Lambda>\frac{1}{r^2}$. This behavior is shown in Fig. 1.
\begin{figure}[h]
\includegraphics[scale=0.7]{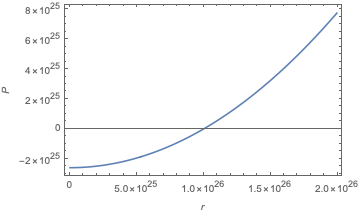}
\caption{Casimir pressure as a function of the distance $r$.}
\end{figure}

\subsection{Gravitational Casimir effect at finite temperature}

Now the topology $\Gamma_4^2=\mathbb{S}^1\times\mathbb{S}^1\times\mathbb{R}^{2}$ with $\alpha=(\beta,i2d,0,0)$ is considered. In this case the double compactification consists in one being the time and the other along the coordinate $r$. The Bogoliubov transformation is
\bea
w^2(\beta,d)=\sum_{l_0=1}^\infty e^{-\beta k^0l_0}+\sum_{l_1=1}^\infty e^{-i2dk^1l_1}+2\sum_{l_0,l_1=1}^\infty e^{-\beta k^0l_0-i2dk^1l_1}.\label{BT3}
\eea
The first two terms are associated with the Stefan-Boltzmann law and the Casimir effect at zero temperature. The Green function for the third term is 
\bea
\overline{G}_0(x-x';\beta,d)&=&4\sum_{l_0,l_1=1}^\infty G_0\left(x-x'-i\beta l_0n_0-2dl_1n_1\right).\label{GF3}
\eea
Proceeding similarly to the previous cases, the Casimir energy at finite temperature is
\bea
E(d,\beta)=-256\sum_{l_o,l_1}^\infty \frac{\left[(2dl_1)^2\left(1+\frac{2r^2}{R^2}\right)-3(\beta l_0)^2\right]}{\left[(2dl_1)^2\left(1+\frac{r^2}{R^2}\right)+(\beta l_0)^2\left(1-\frac{r^2}{R^2}\right)\right]^3}
\eea
and the Casimir pressure at finite temperature is
\bea
P(d,\beta)=-256\sum_{l_o,l_1}^\infty \frac{\left[3(2dl_1)^2-(\beta l_0)^2\left(1-\frac{2r^2}{R^2}\right)\right]}{\left[(2dl_1)^2\left(1+\frac{r^2}{R^2}\right)+(\beta l_0)^2\left(1-\frac{r^2}{R^2}\right)\right]^3}.
\eea
Here, both effects are displayed, i.e., temporal and spatial compactification. In addition, the term $\frac{r^2}{R^2}$ is related to the acceleration $a$. Then, these quantities at finite temperature are modified due to the acceleration of the particle.

\section{Conclusion} \label{sec.5}

Gravity and statistical physics are related by temperature. The Hawking temperature and the Unruh effect are examples of these phenomena. These relations lead to effects associated with quantum physics. In order to satisfy the temperature of gravitation, the TFD formalism, a real-time formalism, is used. The gravitational theory chosen is the teleparallel gravity, which has a well-defined energy-momentum tensor. The de Sitter space-time is considered and then some applications using the gravitational energy-momentum tensor are investigated. The gravitational Stefan-Boltzmann law is calculated. It is shown that the acceleration of the particle in the de Sitter space-time modifies this result. In this context, the gravitational entropy is studied and then the temperature of gravitation in the de Sitter space-time is obtained. It is shown that the temperature is related to the acceleration of the particle. Then it is an Unruh-type effect. Indeed, in the Unruh effect, the temperature gets hotter when the acceleration increases. However, in this case, the temperature cools down with the acceleration. So, our result is an anti-Unruh effect. Another application that has been developed is the Casimir effect at zero temperature. It is displayed that this effect depends on the acceleration. In addition, a phase transition between an attractive and repulsive Casimir effect is considered. This phase transition leads to a critical value for the cosmological constant that defines such a transition between the negative and positive Casimir pressure. Such critical value is perceived by the inflection point in the pressure graph, which is a unique result of gravitational TFD. Furthermore, the gravitational Casimir effect at finite temperature is analyzed. This result exhibits both effects, temporal and spatial compactification. It is also worth noting that the relationship between acceleration and gravity, with the exception of the electromagnetic phenomena, is well established by the principle of equivalence.  It is to be expected that there is a temperature associated with acceleration as there is such an association for gravity in a black hole.  In the case of the de Sitter space there is no restriction on the cosmological constant from the metric point of view.   This means that the cosmological constant can be null without causing a divergence in geometry. On the other hand, the relationship of temperature to acceleration excludes a vanishing cosmological constant, since this would imply a divergent temperature.  With this, we suspect that the so-called dark energy may not only be an expression of the cosmological constant, since it is (if it exists) very small in a Universe whose average temperature is very low.

\section*{Acknowledgments}

This work by A. F. S. is supported by CNPq projects 430194/2018-8 and 313400/2020-2.


\begin{thebibliography}{99}

\bibitem{penrose} R. Penrose and R.M. Floyd, Nature {\bf 229}, 177, (1971).
\bibitem{hawking} S. W. Hawking, Physical Review Letters, {\bf 26}, 1344 (1971).
\bibitem{Bekenstein} J. D. Bekenstein, Physical Review D, {\bf 7}, 2333, (1973).
\bibitem{Fulling} S. A. Fulling, Phys. Rev. D {\bf 7}, 2850 (1973).
\bibitem{Davies} P. C. W. Davies, J. Phys. A. {\bf 8}, 609 (1975).
\bibitem{Unruh} W. G. Unruh, Phys. Rev. D {\bf 14}, 870 (1976).
\bibitem{Crispino} L. C. B. Crispino, A. Higuchi, G. E. A. Matsas, Rev. Mod. Phys. {\bf 80}, 787 (2008).
\bibitem{Gib} G. W. Gibbons and S. W. Hawking, Phys. Rev. D {\bf 15}, 2738 (1977).
\bibitem{Dewitt} B. DeWitt. {\it General Relativity; an Einstein Centenary Survey}, Cambridge University Press, Cambridge, UK, (1980).
\bibitem{Birrel} N. D. Birrell, and P. C. W. Davies. {\it Quantum fields in curved space.} No. 7. Cambridge university press, (1984).
\bibitem{Anti} W. G. Brenna, Robert B. Mann and E. M. Martínez, Phys. Lett. B {\bf 757}, 307 (2016).
\bibitem{Matsubara} T. Matsubara, Prog. Theor. Phys. {\bf 14}, 351 (1955).
\bibitem{Schwinger}J. Schwinger, J. Math. Phys. {\bf 2}, 407 (1961); J. Schwinger, Lecture Notes Of Brandeis University
Summer Institute (1960).
\bibitem{Umezawa1}Y. Takahashi and H. Umezawa, Coll. Phenomena {\bf 2}, 55 (1975); Int. Jour. Mod. Phys. B {\bf 10}, 1755 (1996).
\bibitem{Umezawa2}Y. Takahashi, H. Umezawa and H. Matsumoto, {\it Thermofield Dynamics and Condensed States}, North-Holland, Amsterdan, (1982).
\bibitem{Khanna0} F. C. Khanna, A. P. C. Malbouisson, J. M. C. Malboiusson and A. E. Santana, {\it Themal quantum field theory: Algebraic aspects and applications}, World Scientific, Singapore, (2009).
\bibitem{Umezawa22} H. Umezawa, {\it Advanced Field Theory: Micro, Macro and Thermal Physics}, AIP, New York, (1993).
\bibitem{Khanna1} A. E. Santana and F. C. Khanna, Phys. Lett. A {\bf 203}, 68 (1995).
\bibitem{Khanna2} A. E. Santana, F. C. Khanna, H. Chu, and C. Chang, Ann. Phys. {\bf 249}, 481 (1996).

\bibitem{Acquaviva}
G. Acquaviva, A. Iorio and L. Smaldone, PHYSICAL REVIEW D, {\bf 102}, 106002, (2020).

\bibitem{abdala} M. C. B. Abdalla, A. L. Gadelha and D. L. Nedel, Phys. Lett. B 613, 213 (2005).
\bibitem{Einstein} A. Einstein, Math. Annal. {\bf 102}, 685 (1930).
\bibitem{MalufRev} J. W. Maluf, Ann. Phys. (Berlin) {\bf 525}, 339 (2013).
\bibitem{Maluf} J. W. Maluf, J. Math. Phys. {\bf 37}, 6293 (1996). 
\bibitem{flatTFD}S. C. Ulhoa, A. F. Santos and F. C. Khanna, Gen. Rel. Grav. {\bf 49},  54, (2017).
\bibitem{Casimir} H. G. B. Casimir, Proc. K. Ned. Akad. Wet. {\bf 51}, 793 (1948).
\bibitem{GBT}F. C. Khanna, A. P. C Malbouisson, J. M. C. Malbouisson and A. E. Santana, Ann. Phys. {\bf 326}, 2634 (2011). 
\bibitem{cartan} E. Cartan, On a generalization of the notion of Reimann curvature and spaces with torsion, in NATO ASIB Proc. 58: Cosmology and Gravitation: Spin, Torsion, Rotation, and Supergravity, eds. P. G. Bergmann and V. de Sabbata (1980), pp. 489-491.
\bibitem{Garay}
Luis J. Garay, Eduardo Martín-Martínez and José de Ramón.
Phys. Rev. D, {\bf 94}, 104048, (2016).

\end{thebibliography}
\end{document}